\documentclass[3p,times,procedia]{elsarticle}

\usepackage{ecrc}

\volume{00}

\firstpage{1}

\journalname{Physics Procedia}

\runauth{M. Tabata et al.}

\jid{phpro}

\jnltitlelogo{Physics Procedia}

\CopyrightLine{2011}{}

\usepackage{amssymb}

\usepackage[figuresright]{rotating}

\begin{document}

\begin{frontmatter}



\dochead{TIPP 2011 -  Technology and Instrumentation for Particle Physics 2011}

\title{Recent progress in silica aerogel Cherenkov radiator}


\author[label1,label2]{Makoto Tabata\footnote{Email: makoto@hepburn.s.chiba-u.ac.jp}}
\author[label3]{Ichiro Adachi}
\author[label2]{Hideyuki Kawai}
\author[label2]{\\Masato Kubo}
\author[label2]{Takeshi Sato}

\address[label1]{Institute of Space and Astronautical Science (ISAS), Japan Aerospace Exploration Agency (JAXA), Sagamihara, 252-5210, Japan}
\address[label2]{Department of Physics, Chiba University, Chiba, 263-8522, Japan}
\address[label3]{Institute of Particle and Nuclear Studies (IPNS), High Energy Accelerator Research Organization (KEK), Tsukuba, 305-0801, Japan}

\begin{abstract}
In this paper, we present recent progress in the development of hydrophobic silica aerogel as a Cherenkov radiator. In addition to the conventional method, the recently developed pin-drying method for producing high-refractive-index aerogels with high transparency was studied in detail. Optical qualities and large tile handling for crack-free aerogels were investigated. Sufficient photons were detected from high-performance aerogels in a beam test.
\end{abstract}

\begin{keyword}
Silica aerogel \sep Cherenkov radiator



\end{keyword}

\end{frontmatter}


\section{Introduction}
Silica aerogel is a three-dimensional structural solid of silica particles, and its texture is same as that of styrofoam. It is a highly porous material with low bulk density and is a good thermal insulator. In addition, silica aerogel is a transparent material because it contains a fine structure of silica networks and air on the order of 10 nm; that is, its optical properties are explained by Rayleigh scattering. Aerogel is first synthesized as an alcogel containing alcohol as the solvent in structural pores. The alcohol is then extracted, exchanging it for air by using the supercritical drying method.

Aerogel has been used in high-energy and nuclear physics experiments since the 1970s, because it is the most convenient medium for creating a Cherenkov radiator owing to its unique refractive index. The gap in the refractive index between gas and liquid materials can only be filled by aerogel. The refractive index of aerogel is proportional to its bulk density and can be adjusted in the process of production. The optical transparency of aerogel strongly depends on the conditions during alcogel synthesis. Thus, the development of new techniques for producing aerogel is the key for obtaining the ideal Cherenkov radiator. Therefore, physicists have attempted to customize the method of producing aerogel \cite{cite1,cite2}, although only a few companies (for example, Panasonic Electric Works) manufacture it. The first technique they developed for producing aerogel used a sol-gel single-step method, and high transparency was only available in the refractive index range $n$ = 1.02$�-$1.03. The second technique they developed used a two-step method. Finally, the KEK group developed the KEK method as the third technique, which opened up production of $n = 1.01$ and lower aerogels in the 1990s \cite{cite3}. For example, threshold Cherenkov counters with $n = 1.024$ aerogels were developed for the TASSO experiment at DESY \cite{cite1}, a ring imaging Cherenkov (RICH) detector with $n = 1.03$ was used for the LHCb detector at CERN \cite{cite4}, and threshold Cherenkov counter modules with $n$ = 1.01$�-$1.03 were installed in the Belle spectrometer at KEK \cite{cite5}.

In contrast, the higher-refractive-index range has not yet been fully exploited. Aerogel with $n$ = 1.06$-�$1.08 can be directly synthesized but the transparency is far from satisfactory. The sintering of manufactured aerogel is a known method for producing high-refractive-index samples with up to $n = 1.20$. Only one usage of a high-refractive-index aerogel has been reported: the use of an $n = 1.13$ aerogel for the SND detector upgrade at BINP \cite{cite6}.

At the previous TIPP09 conference, we reported the successful production of hydrophobic aerogel with a wide refractive index range ($1.0026 < n < 1.26$) \cite{cite7}. Since then, we have developed highly transparent aerogels with a high refractive index by a novel production method: the pin-drying method. In this paper, we report the development and manufacturing status of aerogel Cherenkov radiators by both the pin-drying and conventional methods.

\section{Production methods}
The pin-drying method is a technique for generating alcogels with densities higher than that of the starting alcogel while maintaining the original transparency. In the TIPP09 conference, we called it the ``pinhole drying'' method. The name ``pin-drying'' method derives from its introduction in the conference summary talk \cite{cite8}. The essence of the pin-drying method is shrinking alcogels while maintaining the aspect ratio and preventing cracking. It is achieved by using a semi-sealed pin container to partially evaporate the solvent contained in the alcogels. The pin-drying method is based on our conventional method of synthesizing the starting alcogel. The process after pin-drying of the starting alcogel is also identical to that of the conventional method.

Our conventional method of producing aerogels is based on the KEK method (the third production method listed in the Introduction). The original KEK method specialized in producing aerogels with low refractive index ($n$ = 1.01$�-$1.03); however, it has been extended to obtain transparent aerogels with higher refractive index ($n$ = 1.04$�-$1.07). The novel solvent \textit{N},\textit{N}-dimethylformamide (DMF) has been introduced in the alcogel synthesis process \cite{cite9}. Ethanol, methanol, and DMF are used as different solvents according to the desired refractive index. Here we call the generalized wet-gel a ``solvogel'' instead of an alcogel.

\begin{figure}[t] 
\centering 
\includegraphics[width=0.378\textwidth,keepaspectratio]{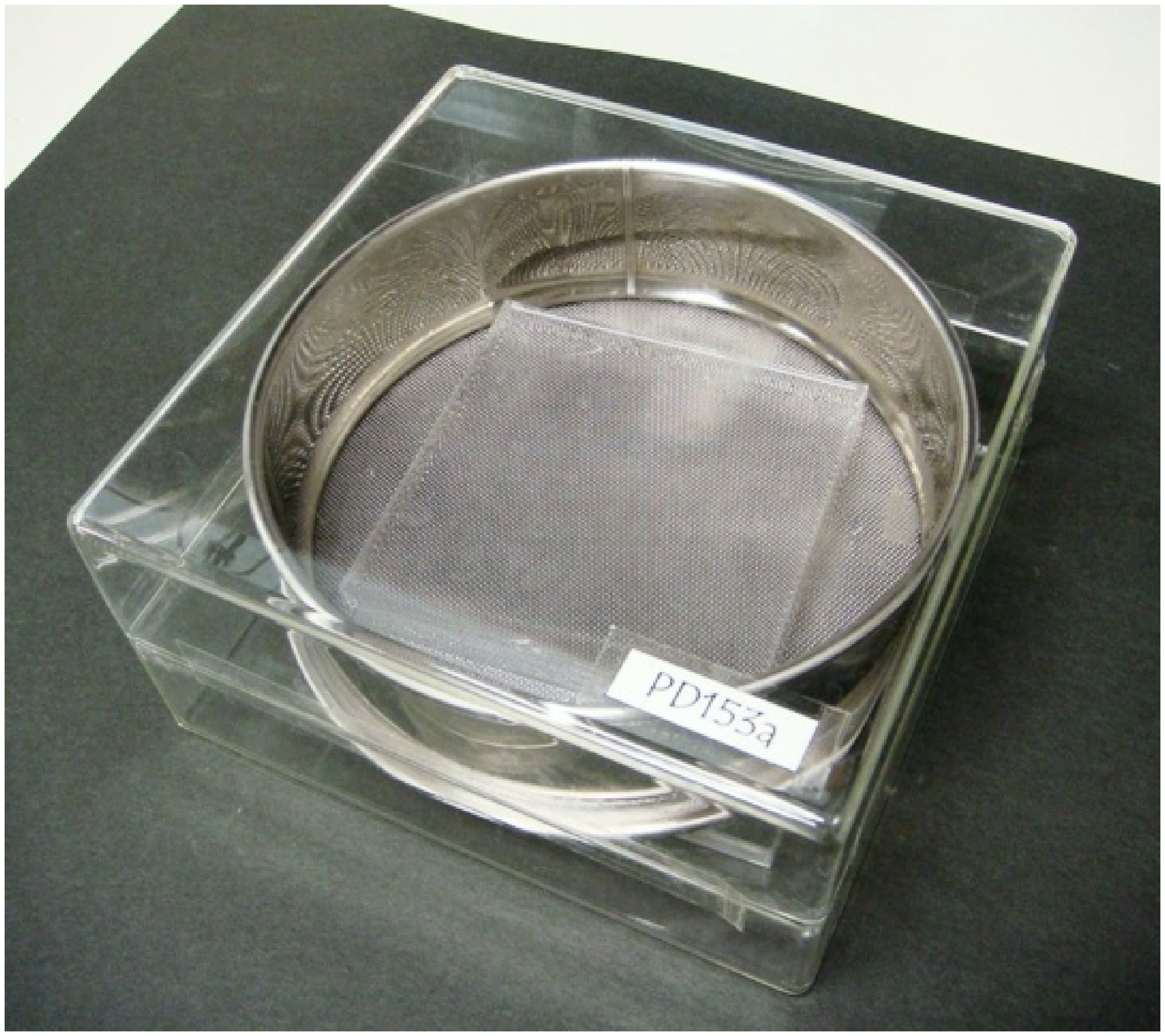}
\includegraphics[width=0.378\textwidth,keepaspectratio]{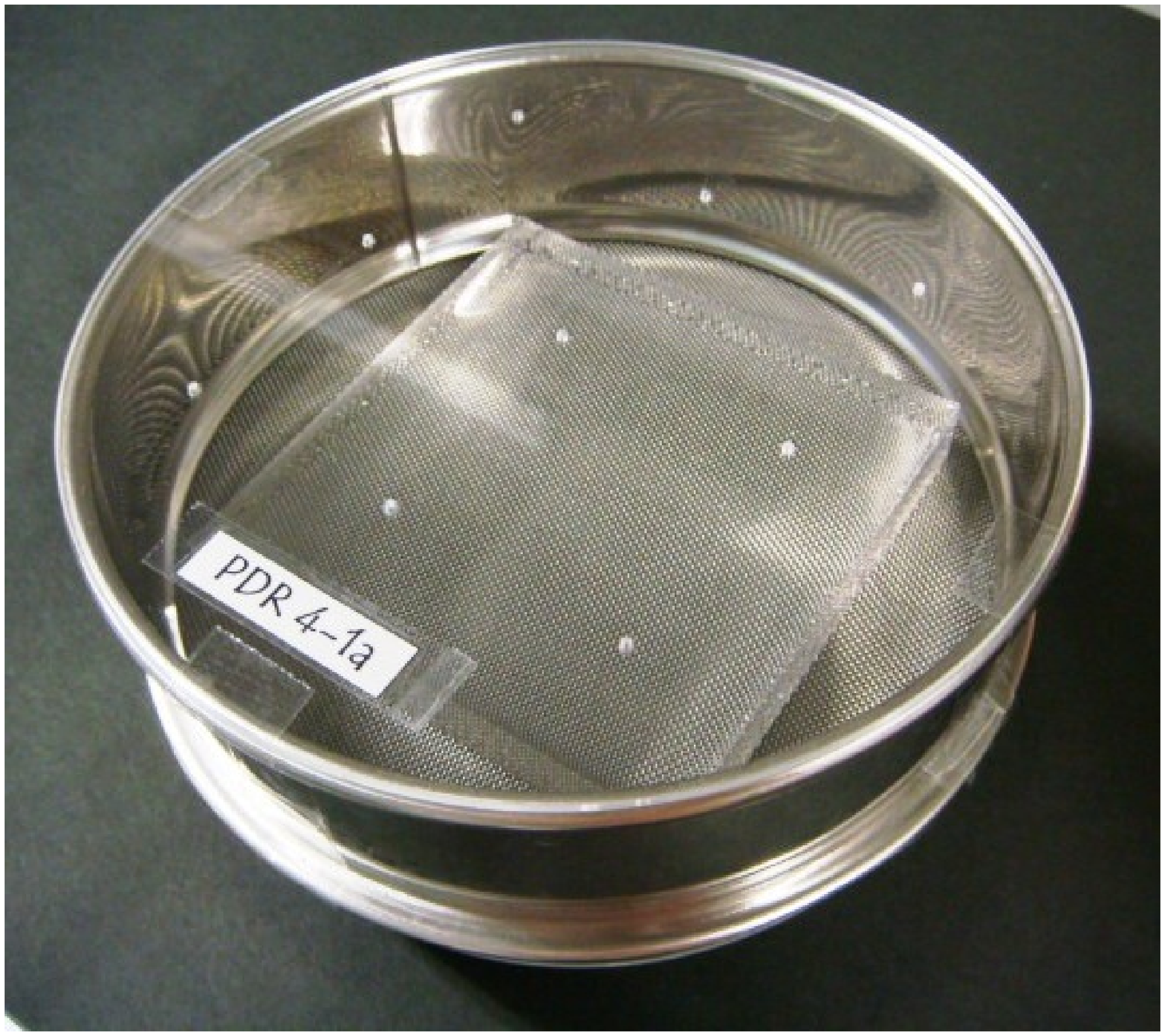}
\label{fig:fig1}
\caption{Two types of pin containers: type-g (left) and type-a (right).}
\end{figure}

\subsection{Conventional production method}
Aerogel production, which takes about one month, begins by the synthesis of a solvogel followed by one week aging in a sealed polystyrene mold. The solvogel production is completed here. After the solvogel is immersed in ethanol and detached from the mold to exchange the solvent for ethanol, hydrophobic treatment is performed in ethanol for three days. Extra ammonia generated in the hydrophobic process is removed by replacing ethanol three times. Finally, supercritical drying by liquefied carbon dioxide is performed using an autoclave. The complete production procedure is detailed in Ref. \cite{cite10}. Note that the solvogel always remains in ethanol or the sealed mold during conventional production.

\begin{figure}[t] 
\centering 
\includegraphics[width=0.35\textwidth,keepaspectratio]{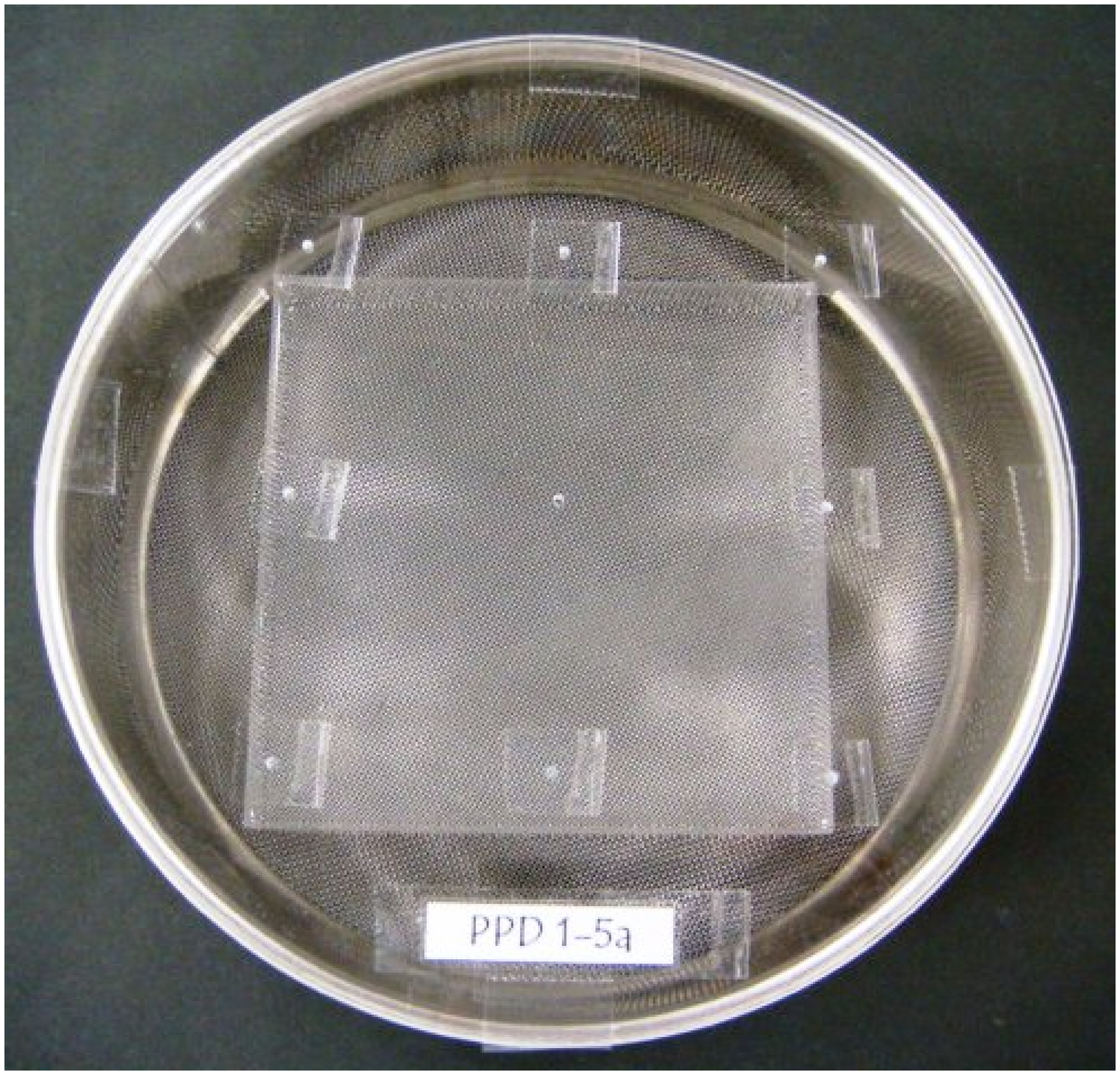}
\includegraphics[width=0.35\textwidth,keepaspectratio]{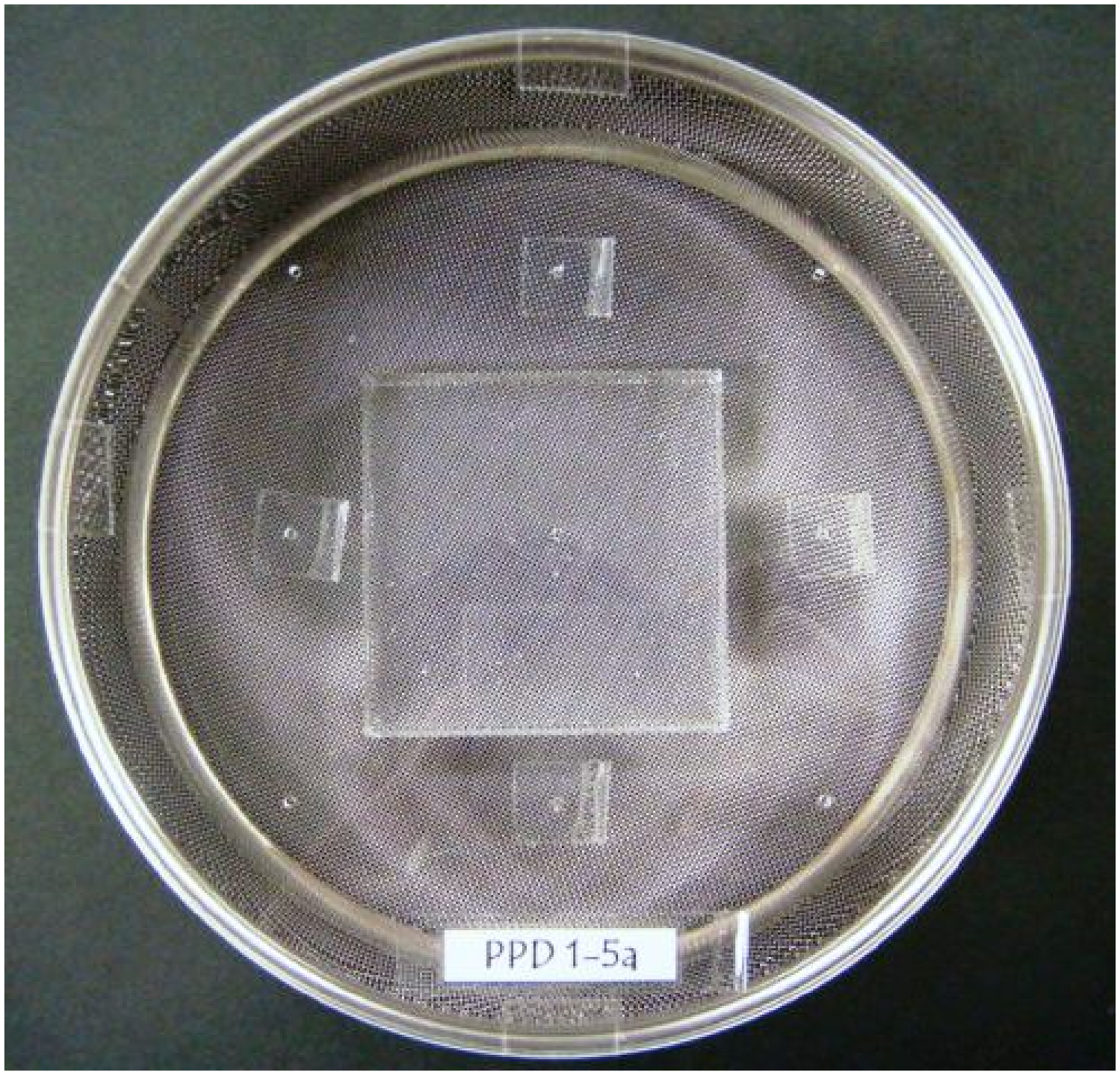}
\label{fig:fig2}
\caption{Photographs of a solvogel before (left) and after (right) the pin-drying process.}
\end{figure}

\subsection{Pin-drying production method}
In the pin-drying method, the pin-drying process is inserted between the aging and hydrophobic treatment processes. First, a solvogel with $n$ = 1.045$�-$1.066 is synthesized using methanol or DMF as the solvent. After only three days of aging, the completed solvogel is rapidly detached from the mold in the ethanol bath in a process which ensures that the solvent used for solvogel synthesis is not substituted for ethanol. Then, the solvogel is carefully positioned meniscus side up on a strainer with 500-$\mu $m pitch in ethanol. The strainer is chosen among $\phi $15, 20, and 30 cm diameters according to the solvogel size. The strainer retrieved from the ethanol bath is semi-sealed to create a pin container.

Two types of pin containers were designed, as shown in Fig. 1: type-g and type-a. In the type-g pin container, the strainer is sandwiched between two glass cases with natural voids in the joint part, because the glass case is handmade. The two glass cases are clamped with scotch tape, which functions as a void adjuster. The glass case was originally made as a mold for synthesizing low-density alcogels, but it was diverted for the $\phi $15 cm strainer. The type-g pin container is relatively stable and reliable for pin-drying production. In the type-a pin container, a large strainer is sandwiched between two acrylic plates with a thickness of 2 mm and with some pinholes. In this container as well, scotch tape is used to attach the strainer and acrylic plates. If the interface between the strainer and the acrylic plates is not completely filled, it also functions like pinholes.

Fig. 2 shows the shrinkage of a solvogel in a pin container (type-a). It takes several weeks or months to complete the pin-drying process, depending on the conditions of the starting solvogel and the target refractive index. The number of pinholes and the width of voids should be controlled throughout the process to avoid cracking of the solvogel. To determine the end of the pin-drying process, the weight of the solvogel is indirectly monitored. The pin container is not opened until the end of the process.

After the pin-drying process, hydrophobic treatment, identical to that performed in the conventional method, is performed. For this purpose, the solvogel is immersed in ethanol in a recovery process. For the case of a target with $n$ = 1.050$�-$1.075, the pin container is directly sunk into the ethanol bath and the solvogel is retrieved. However, the recovery process for a target with $n > 1.10$ is somewhat complicated. To avoid cracking, the pin container is first sealed in an air-tight pod filled with ethanol vapor. In a few days, ethanol vapors penetrate the pin container to moisturize the solvogel. A small amount of ethanol is then added to the pin container to accelerate moisturization, and finally, the pin container is sunk into ethanol. Moreover, hydrophobic treatment should be conducted step-by-step to avoid cracking for $n > 1.10$. Ammonia removal is performed followed by supercritical drying, identical to that performed in the conventional method. The pin-drying process is never a substitution for supercritical drying.

In the single- and two-step methods and the KEK method, the refractive index of aerogels is determined by the preparation recipe for starting chemicals in the sol-gel step. The pin-drying process provides one more opportunity to control the refractive index. By controlling the refractive index at two stages, the pin-drying method enables the creation of aerogels with ultrahigh refractive index and sufficient transparency. Because the pin-drying method allows two-stage control of the refractive index, we consider it to be the fourth method for producing aerogels.

\section{Optical performance}
\label{sec:sec3}

\begin{figure}[t] 
\centering 
\includegraphics[width=0.70\textwidth,keepaspectratio]{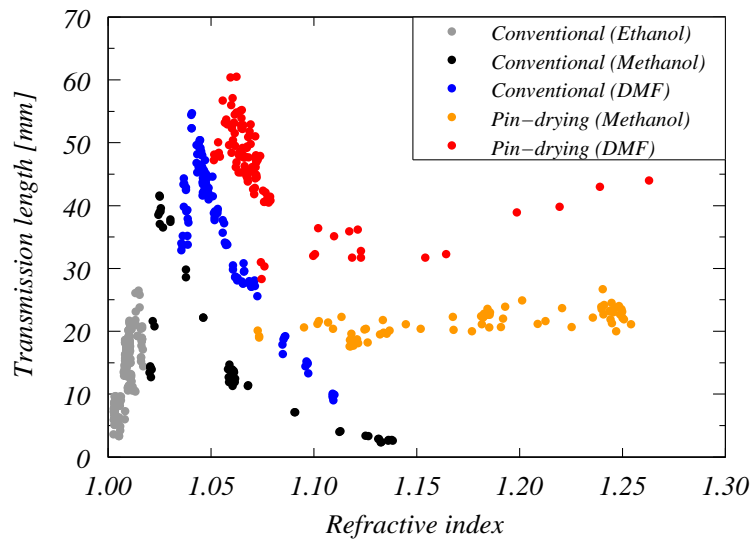}
\label{fig:fig3}
\caption{Transmission length at 400 nm as a function of the refractive index at 405 nm. Each data point represents a different aerogel. Aerogels produced by the conventional method are classified by the solvent used during their synthesis: ethanol (gray), methanol (black), and DMF (blue). Aerogels produced by the pin-drying method are also classified by the solvent used: methanol (orange) and DMF (red).}
\end{figure}

Fig. \ref{fig:fig3} shows transmission length at 400 nm as a function of the refractive index for typical aerogel samples manufactured at Chiba University. The transmission length $\Lambda _T$ is defined as $\Lambda _T(\lambda ) = -t/{\rm ln}T(\lambda )$, where $\lambda $ is the wavelength of light, $t$ is the thickness of the aerogel samples, and $T$ is the transmittance measured with a spectrophotometer. Detailed methods for measuring the optical parameters are given in Ref. \cite{cite10}.

Fig. \ref{fig:fig3} consists of samples manufactured by both the conventional and pin-drying methods. The samples produced by the conventional method are further classified by the solvent used for solvogel synthesis: ethanol, methanol, and DMF. For the alcohol solvents, the most transparent sample ($\Lambda _T$ = 40 mm) was obtained at $n = 1.025$; both above and below this refractive index, the transmission length decreased rapidly. Although aerogels with refractive indices up to $n = 1.14$ can be produced by the conventional method with methanol, transmission lengths longer than 10 mm only occurred up to $n = 1.07$. As a result of introducing DMF as a solvent in the conventional method, a significant improvement in the transmission length was attained in the range $n$ = 1.035$�-$1.11; in particular, it was remarkable in the range $n$ = 1.04$�-$1.07, and the transmission length reached over 50 mm at $n = 1.04$.

The samples produced by the pin-drying method exhibited even better optical performances. First, solvogels with a starting refractive index of $n_s = 1.06$ synthesized with methanol solvent were studied. This resulted in aerogels with ultrahigh refractive indices of $n$ = 1.07$-�$1.25 with sufficient transparency ($\Lambda _T >$ 20 mm) and with no cracking.

Second, solvogels with a starting refractive index of $n_s = 1.066$ synthesized with DMF solvent were studied. This resulted in aerogels with ultrahigh refractive indices of $n$ = 1.075$-�$1.26 with superior transparency ($\Lambda _T >$ 30 mm). However, note that it took longer to complete the pin-drying process with DMF because the evaporation rate of DMF is lower. Furthermore, it was difficult to obtain crack-free samples in the range $n > 1.20$.

In any case, pin-dried aerogels exhibited better transparencies than the starting aerogels. This was applied to aerogels with middle refractive indices in the range $n$ = 1.05$-�$1.08. Solvogels with $n_s = 1.049$ were chosen as the starting solvogels. As shown in Fig. \ref{fig:fig3}, the longest transmission length ($\Lambda _T$ = 60 mm) was achieved at $n \sim  1.06$. In the range $n$ = 1.05$-�$1.08, a transmission length above 40 mm was recorded for the pin-dried aerogels created using DMF as the solvent.

\section{Productivity in the pin-drying method}
\label{sec:sec4}

\begin{figure}[t] 
\begin{minipage}[t]{0.48\textwidth}
\centering 
\includegraphics[width=1\textwidth,keepaspectratio]{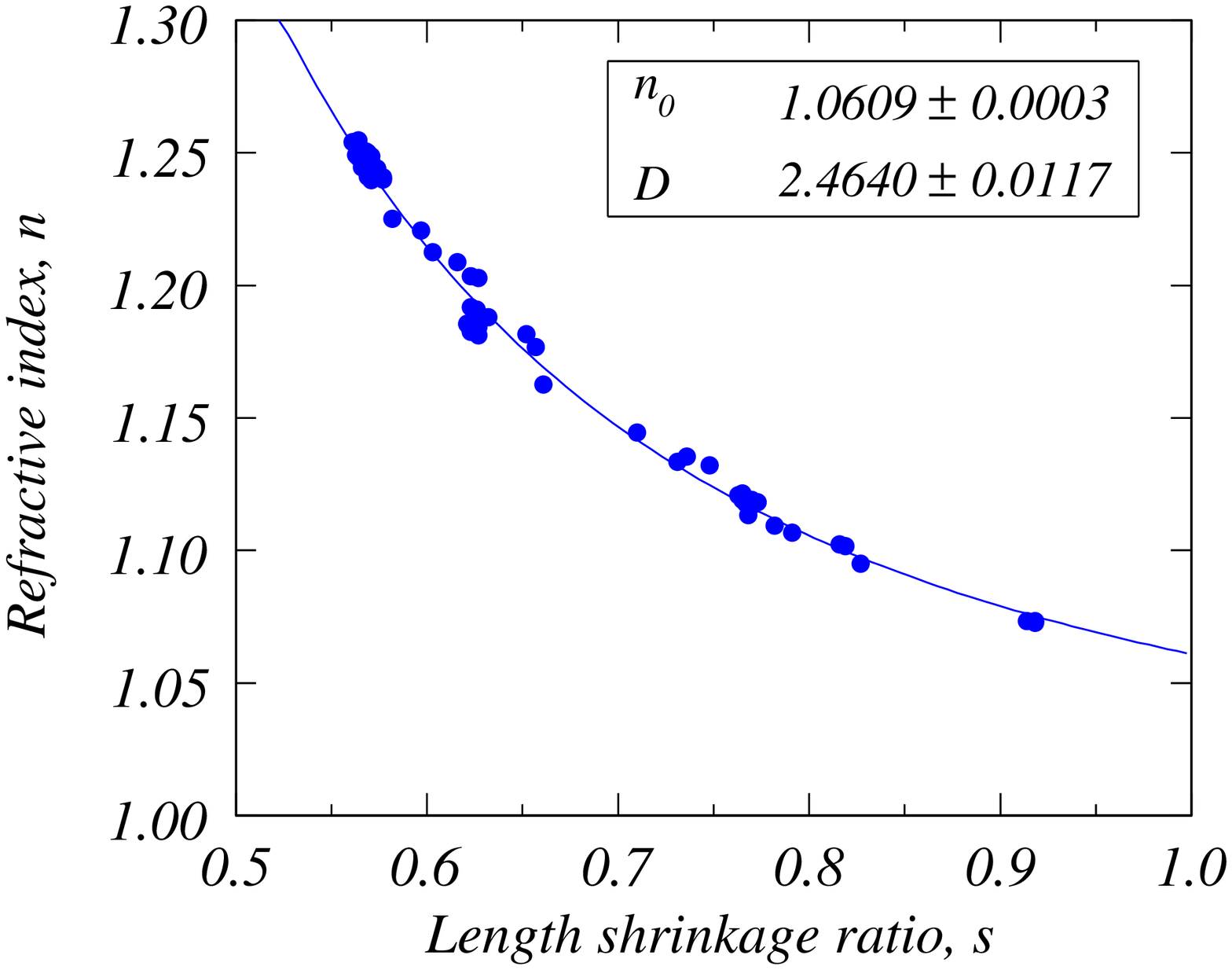}
\label{fig:fig4}
\caption{Final refractive index $n$ at 405 nm as a function of the shrinkage ratio in length $s$ for the 66 final aerogels. The parameter values of the fitting function $n = 1 + (n_0-1) s^{-D}$ are shown on the plot.}
\end{minipage}
\hfill
\begin{minipage}[t]{0.48\textwidth}
\centering 
\includegraphics[width=1\textwidth,keepaspectratio]{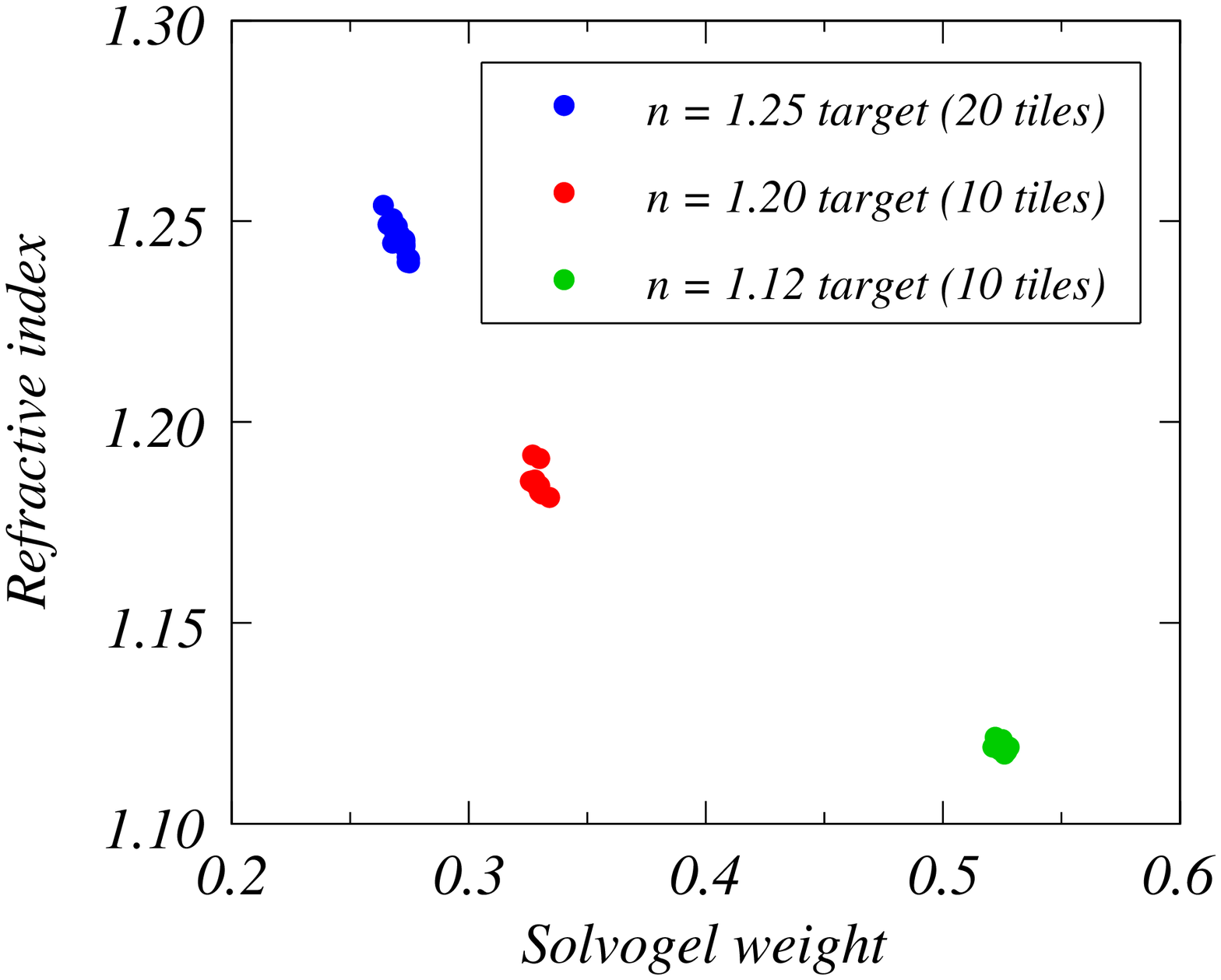}
\label{}
\caption{Final refractive index at 405 nm as a function of the relative weight for 40 solvogels. The solvogel weight is normalized by the initial solvogel weight in the pin-drying process.}
\end{minipage}
\end{figure}

Further details of the pin-drying process depend on conditions such as the starting refractive index, the solvent used for solvogel synthesis, and the size of the solvogels. In this section, the results of several trial aerogel productions are given as an example. The experimental conditions were as follows:
\begin{itemize}
 \item Starting refractive index: $n_s \sim  1.060$
 \item Solvent for solvogel synthesis: methanol
 \item Dimensions of starting solvogels (mold size): 96 $\times $ 96 mm$^2$ or 120 $\times $ 82 mm$^2$
 \item Thickness of final aerogels: 10 mm
 \item Pin container: type-g
\end{itemize}

Fig. \ref{fig:fig4} shows the achieved refractive index as a function of the length shrinkage ratio defined as $s = l/l_0$, where $l$ and $l_0$ are the length of the longest side of the final aerogels and that of the molds for solvogel synthesis, respectively. The solid line represents the best fit of a function to all the aerogel data points. The fitting function is defined as $n = 1 + (n_0-1) s^{-D}$, where $n_0$ and $D$ are parameters; as a result of the fitting, the parameters were given as $n_0 = 1.0609 \pm  0.0003$ and $D = 2.464 \pm  0.012$. On the basis of the fitting curve, the final refractive index can be roughly estimated by measuring the length of the solvogel in the pin-drying process. Here shrunken solvogels are assumed not to expand during the subsequent processes.

However, it is difficult to accurately measure the length of solvogels from outside the pin containers. Instead, the weight of the solvogels was indirectly monitored during the pin-drying process. To judge the end of the pin-drying process from the solvogel weight, the weight of the pin containers must be known, and the solvogel weight must be related to the final refractive index in advance. The relationship between solvogel weight and final refractive index is shown in Fig. 5. Three trial aerogel productions to target $n = 1.25$, 1.20, and 1.12 are demonstrated here. The solvogel weight is normalized by the initial solvogel weight in the pin-drying process. Because of the lack of tests in advance, the $n = 1.20$ target aerogels shifted toward lower refractive indices. In general, the final refractive index can be well controlled by monitoring the solvogel weight. The full widths of the refractive index variation were 0.014, 0.011, and 0.004 for the $n = 1.25$, 1.20, and 1.12 targets, respectively. The durations of the pin-drying were 32$�-$57 days at 23$^\circ $C, 26$-�$33 days at 22$^\circ $C, and 15$-$18 days at 27$^\circ $C for the $n = 1.25$, 1.20, and 1.12 targets, respectively.

\section{Large tile production}
Large aerogel tile production with no cracking is another key factor in obtaining high performance, because it reduces radiator boundaries in a large area of a RICH counter. The larger the volume and the higher the density of an aerogel, the more difficulty it has in completing supercritical drying without cracking. In 2005, typical dimensions of our aerogel tiles were 11 $\times $ 11 $\times $ 2 cm$^3$ in the range $n$ = 1.05$-�$1.06. Since then, we have managed to produce 15 $\times $ 15 $\times $ 2 cm$^3$ tiles with no cracking in the same refractive index range. Recently, we ascertained that it is important to slowly ramp down the pressure to atmospheric pressure during the supercritical drying process to reduce cracking. By using a custom-made polystyrene mold, large-area (18 $\times $ 18 $\times $ 2 cm$^3$) aerogel tiles with $n \sim  1.05$ were manufactured in 2011 with no cracking, good yield, and high transparency by the conventional method, as shown in Fig. 6.

However, further careful handling of solvogels is needed during the pin-drying production method because the solvogel is taken out of ethanol. In initial trial productions, starting solvogels with dimensions of 9 $\times $ 9 $\times $ 2 cm$^3$ were studied. A later large tile production study successfully produced final aerogels with dimensions of 14 $\times $ 14 $\times $ 2 cm$^3$ and with $n \sim  1.06$. For aerogels with ultrahigh refractive indices, aerogels with dimensions of 8 $\times $ 8 $\times $ 1 cm$^3$ and 6 $\times $ 6 $\times $ 1 cm$^3$ were produced for $n = 1.10$ and 1.20, respectively, because a large volume reduction is required in the pin-drying process.

\begin{figure}[t] 
\begin{minipage}[t]{0.48\textwidth}
\centering 
\includegraphics[width=1\textwidth,keepaspectratio]{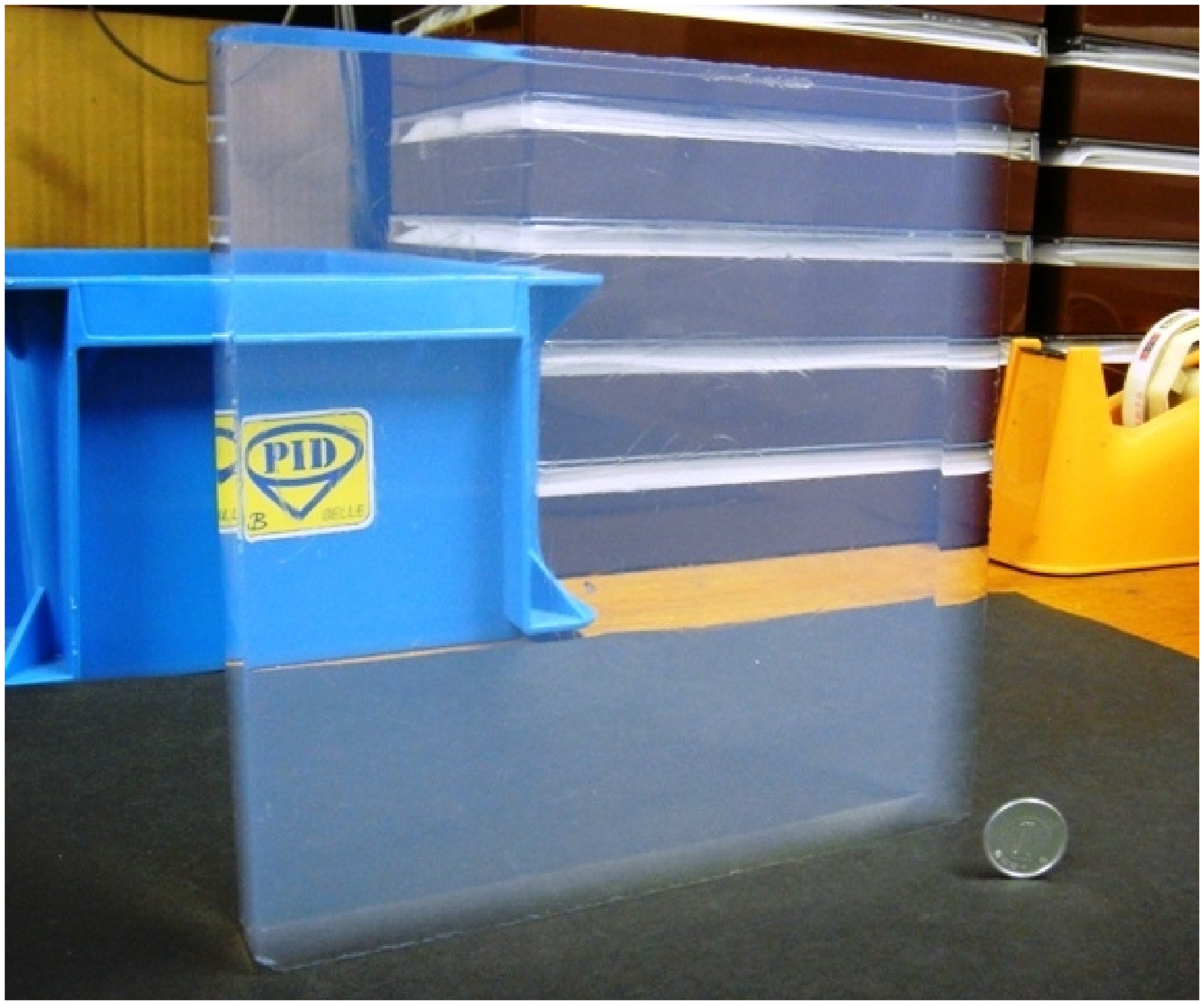}
\label{fig:fig6}
\caption{Photograph of a large aerogel tile with $n = 1.051$ and $\Lambda _T$ = 40 mm. As a reference, the coin has a diameter of 2 cm.}
\end{minipage}
\hfill
\begin{minipage}[t]{0.48\textwidth}
\centering 
\includegraphics[width=1\textwidth,keepaspectratio]{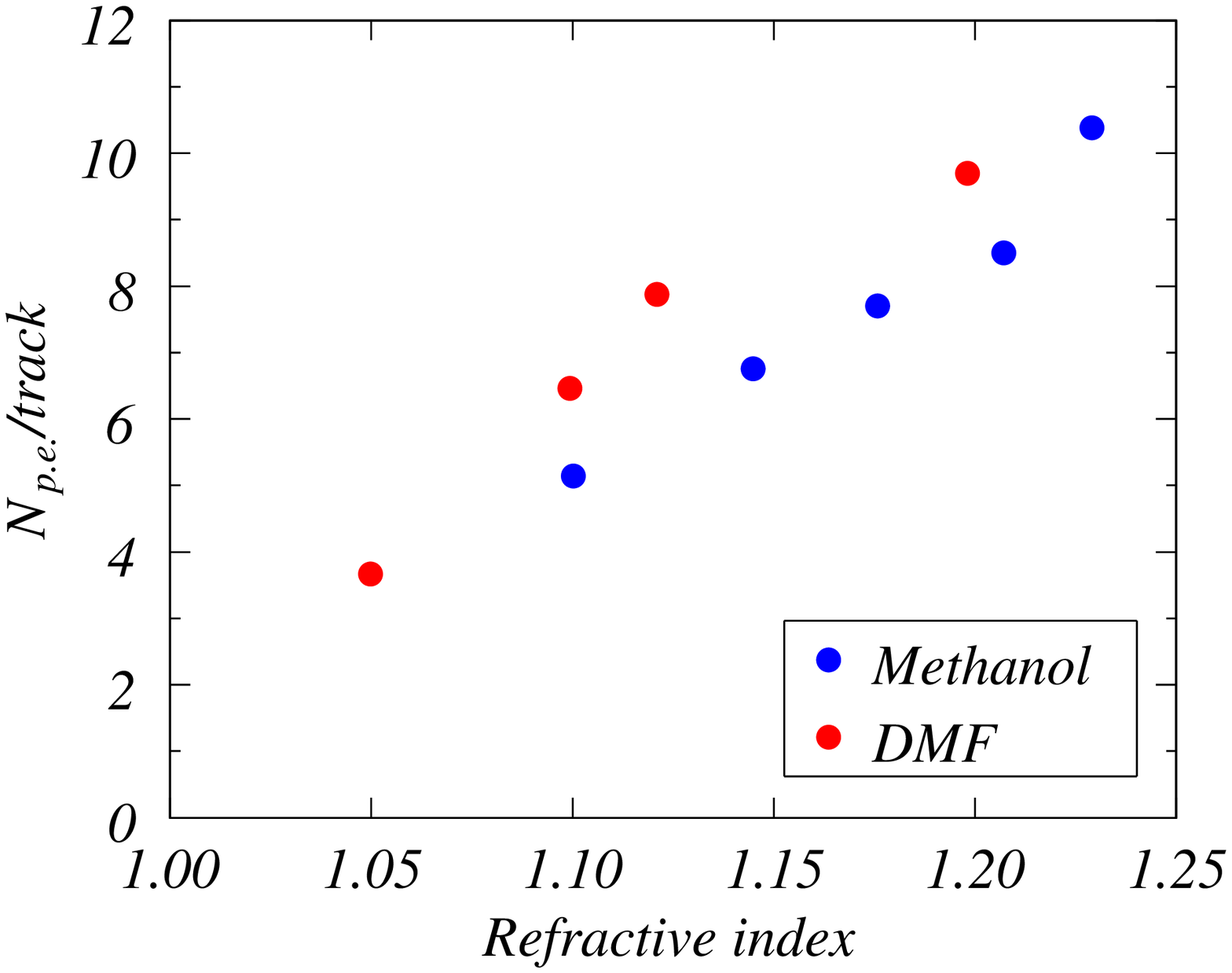}
\label{fig:fig7}
\caption{Number of detected photoelectrons per incident electron track as a function of the refractive index measured at 405 nm. The number of detected photoelectrons is normalized by the aerogel thickness (1 cm). The red and blue points represent aerogels synthesized with DMF and methanol, respectively. An aerogel with $n = 1.05$ was manufactured by the conventional method as a reference, and the other aerogels were produced by the pin-drying method.}
\end{minipage}
\end{figure}

\section{Beam test}
In this section, we present the number of detected photoelectrons $N_{p.e.}$ as the result of a beam test for newly developed aerogel radiators. The beam test experiment was conducted to investigate the performance of a prototype aerogel RICH counter for the Belle II experiment \cite{cite11} in November 2009 at the Fuji test beam line (FTBL) in KEK, where 2 GeV/$c$ electron beams were available. In a light-shielded box, a proximity RICH detector was composed of aerogel radiators and a 2 $\times $ 3 array of 144-channel hybrid avalanche photo-detectors (HAPD) \cite{cite12}, which were read out by ASIC/FPGAs \cite{cite13}. Electron tracks were measured by two multi-wire proportional chambers at the upstream and downstream ends of the light-shielded box. The mean quantum efficiency of the six HAPDs was 24\% at 400 nm. The acceptance of the Cherenkov ring ranged from 50\% to 60\% because the HAPDs were arranged with gaps between their packages.

As a candidate configuration for the Belle II aerogel RICH counter based on the focusing radiator scheme \cite{cite14}, we tested a combination of aerogels with $n = 1.05$ for the upstream layer and $n = 1.06$ for the downstream layer. Each aerogel had a thickness of 2 cm. Aerogels formed by the conventional production method had transmission lengths of 48 and 31 mm for the upstream and downstream layers, respectively. In contrast, aerogels formed by the pin-drying production method had transmission lengths of 48 and 55 mm for the upstream and downstream layers, respectively. The distance between the upstream face of the radiator and the photo-detection plane was set to 20 cm, and the averaged ring acceptance was estimated to be 60\%. In the analysis of the beam test data, we found that $N_{p.e.}$ per incident electron track was 10.6 and 13.6 for the conventional and pin-dried aerogels, respectively. In particular, the improved transparency of the downstream aerogel caused the photoelectron yield to increase by 28\%.

As an additional study, we tested ultrahigh-refractive-index ($n$ = 1.10$-�$1.23) aerogels with a thickness of 1 cm. These aerogels were produced by the pin-drying method using methanol or DMF as the solvent. Depending on their refractive indices, the distances between the radiators and the HAPDs were set such that the radius of the Cherenkov rings was 65 mm on the photo-detection plane. The averaged ring acceptance was estimated to be 51\%. Fig. 7 shows $N_{p.e.}$ per track as a function of the refractive index. We quantitatively counted photons emitted by ultrahigh-refractive-index aerogels by using a RICH counter; this is the first time such a measurement has been performed. We found that the photoelectron yield increased with the refractive index. Comparing aerogels synthesized with DMF with those synthesized with methanol at the same refractive index, aerogels synthesized with DMF are more transparent and thus show a higher photoelectron yield.

\section{Discussion}
For the Belle II experiment at SuperKEKB, the KEK group (including the current authors) is developing a proximity-focusing aerogel ring imaging Cherenkov (A-RICH) detector to separate kaons from pions at 1$�-$4 GeV/$c$ in the forward end cap. To detect clear Cherenkov rings with sufficient photons for both kaons and pions, transparent aerogels are needed in the high-refractive-index range. For this purpose, the pin-drying method of producing aerogels with $n$ = 1.05$�-$1.07 was studied in detail. On the basis of the results shown in Sec. \ref{sec:sec3}, we are planning to use the most transparent aerogels with $n$ = 1.05$�-$1.06 for the A-RICH detector. Based on the issue of cost for the pin-drying method, aerogels for the upstream ($n = 1.05$) and downstream ($n = 1.06$) layers will be produced by the conventional and pin-drying methods, respectively. To cover the large area of the end cap (3.5 m$^2$ $\times $ two layers), large aerogel tile production is important. Aerogels will be trimmed with a water jet cutter for tiling. The cutting surface of aerogels significantly scatters photons. Developing large aerogel tiles reduces the boundaries between tiles; therefore, it has a positive impact on the performance of the A-RICH detector.

In addition, many upcoming nuclear physics experiments that have been proposed and approved at J-PARC indicate a high demand to implement threshold Cherenkov counters using aerogels as radiators for triggering or for particle identification; some of these experiments are already in the detailed-design stage. For example, E03 (measurement of X-rays from $\Xi ^-$ atoms) requires an $n = 1.12$ Cherenkov counter to trigger positive kaons from protons at 1$�-$2 GeV/$c$ \cite{cite15}, and the E16 upgrade plan (exploration of the chiral symmetry in QCD) shows the importance of Cherenkov counters with $n = 1.034$ for a kaon spectrometer. Furthermore, E27 (search for the $K^-pp$ state) also requires $n = 1.25$ to separate kaons from high-momentum protons. For E03, aerogels with $n = 1.12$ have already been manufactured by the pin-drying method using DMF as the solvent, and they are to be installed in a counter. For E27, aerogels with $n = 1.25$ have also been manufactured by the same method using methanol as the solvent as a test before mass production as described in Sec. \ref{sec:sec4}. For the E16 upgrade, aerogels with $n \sim  1.034$ are under trial production by the conventional method using DMF as the solvent. Although transparency in aerogels has been somewhat improved by recent developments in this refractive index range (Fig. \ref{fig:fig3}), high-transparency aerogels should be produced by the pin-drying method in the near future.

\section{Conclusion}
Both the conventional and pin-drying methods of producing hydrophobic silica aerogels with high refractive index ($n < 1.26$) were studied in detail. The refractive index was well controlled in the pin-drying process, and the longest transmission length ($\Lambda _T$ = 60 mm) was recorded at $n \sim  1.06$. In a beam test using the RICH counter, 13.6 photoelectrons were detected from the pin-dried aerogels with 4-cm thickness and $n$ = 1.05$-�$1.06. In addition, up to 10.4 photoelectrons were detected from the pin-dried aerogels with 1-cm thickness and $n < 1.23$. In the conventional method, large aerogel tiles (18 $\times $ 18 $\times $ 2 cm$^3$) were successfully manufactured with high transparency ($\Lambda _T >$ 40 mm) and high refractive index ($n \sim  1.05$) by carefully controlling the pressure reduction process during supercritical drying. This significant progress will open up many opportunities to employ aerogels in Cherenkov counters.

\section*{Acknowledgments}
\label{}
The authors are grateful to the members of the Belle II A-RICH group and Dr. H. Yokogawa of Panasonic Electric Works for their assistance in aerogel development. This study was partially supported by a Grant-in-Aid for Scientific Research on Innovative Areas (No. 21105005) from the Ministry of Education, Culture, Sports, Science and Technology (MEXT).

%



\end{document}